\documentclass[11pt]{article}
\usepackage{moriond,epsfig}

\begin{document}
\vspace*{4cm}
\title{NEUTRINO MASS AND MIXING PARAMETERS: A SHORT REVIEW}

\author{G.L.\ Fogli, {E.\ Lisi}~\footnote{Speaker. 
E-mail: {\tt eligio.lisi@ba.infn.it} },
A.\ Marrone, A.\ Palazzo, A.M.\ Rotunno}

\address{Dipartimento di Fisica and Sezione INFN di Bari, \\
Via Amendola 173, 70126 Bari, Italy}

\maketitle\abstracts{We present a brief
review of the current status of
neutrino mass and mixing parameters, based on a 
comprehensive phenomenological analysis  of
neutrino oscillation and non-oscillation
searches, within the standard three-neutrino mixing
framework.}

\section{Introduction}

There is compelling 
experimental evidence \cite{PDG4} that the three known neutrino states with
definite flavor $\nu_\alpha$ ($\alpha=e,\mu,\tau$)  are linear
combinations of states with definite mass $\nu_i$ ($i=1,2,3$), and
that the Hamiltonian of neutrino propagation in vacuum \cite{Pont} and
matter \cite{Matt} does not commute with flavor. The evidence
for flavor nonconservation (i.e., ``neutrino oscillations'') 
comes from a series of experiments performed
during about four decades of research with very different neutrino
beams and detection techniques: the solar neutrino 
\cite{Ba89} experiments
Homestake \cite{Home}, Kamiokande \cite{Kami}, 
SAGE \cite{SAGE}, GALLEX-GNO \cite{GALL,GGNO},
Super-Kamiokande (SK) \cite{SK04} 
and the Sudbury Neutrino Observatory (SNO)~\cite{SNO2,SNOL,Mikn}; 
the long-baseline reactor neutrino 
experiment KamLAND \cite{Kam2,Berg};
the atmospheric neutrino  
experiments
Super-Kamiokande \cite{AtSK,Sula}, 
MACRO \cite{MACR}, and Soudan-2 \cite{Soud}; and the
long-baseline accelerator neutrino  experiment KEK-to-Kamioka (K2K)
\cite{K2K2,Mari}.

Together with the null results from the CHOOZ
\cite{CHOO} 
short-baseline reactor experiment, the above
oscillation data
provide stringent constraints on the neutrino
mixing matrix, on the splittings between squared neutrino masses,
and on  matter effects. The absolute neutrino masses are
being probed by different, non-oscillation searches: 
beta decay experiments \cite{Eite},
neutrinoless double beta decay searches
($0\nu2\beta$) \cite{Elli,Cape}, and precision cosmology
\cite{Hu98,Be03,Tg04,Selj,Past}. Current non-oscillation data 
provide only upper limits on neutrino masses,
with the only exception of the Heidelberg-Moscow 
$0\nu2\beta$ experiment
\cite{Kl04}, 
whose claimed signal implies a lower bound on neutrino masses.

Basically all the data are consistent with the simplest
extension of the standard electroweak model needed to accommodate
nonzero neutrino masses and mixings, namely, with a scenario where the
three known flavor states $\nu_{e,\mu,\tau}$ are mixed with only three
mass states {$\nu_{1,2,3}$}, no other states or
new neutrino interactions being needed. This
``standard three-neutrino framework''
(as recently reviewed, e.g., in 
\cite{Ours}, where further references can be found) 
appears thus as a new paradigm of particle
and astroparticle physics, which will be tested, refined, and possibly
challenged by a series of new, more sensitive experiments planned
for the next few years or even for the next decades \cite{Futu}. 
The first
challenge might actually come very soon from the running MiniBooNE
experiment \cite{Boon}, which is probing the only piece of data at variance with the
standard three-neutrino framework, namely, the controversial result of the
Liquid Scintillator Neutrino Experiment (LSND) \cite{LSND}.

In this short review we focus on the current status of the standard three-neutrino
framework and on the neutrino mass and mixing parameters which characterize
it. The parameters are defined as follows. The unitary mixing matrix $U$,
in terms of one-particle neutrino states
$|\nu\rangle$,  is defined as (see, e.g., \cite{PDG4}),
\begin{equation}
\label{Ustar}
|\nu_{\alpha}\rangle=\sum_{i=1}^3 U_{\alpha i}^* |\nu_{i}\rangle\ .
\end{equation}
A common parameterization for the matrix $U$ is:
\begin{equation}
\label{Euler}
U=O_{23}\Gamma_\delta O_{13} \Gamma_\delta^\dagger O_{12}\ ,
\end{equation}
where the $O_{ij}$'s are real Euler rotations with angles
$\theta_{ij}\in [0,\pi/2]$, while
$\Gamma_\delta$ embeds a CP-violating phase $\delta\in [0,2\pi]$,
\begin{equation}
\label{Gamma}
\Gamma_\delta=\mathrm{diag}(1,1,e^{+i\delta})\ .
\end{equation}
By considering $\Gamma_\delta O_{13}\Gamma_\delta^\dagger$ as a
single (complex) rotation, this parametrization coincides with the
one recommended [together with Eq.~(\ref{Ustar})] in the Review of
Particle Properties \cite{PDG4}.

For the sake of simplicity,
the phase $\delta$ will not be considered
in full generality hereafter. Numerical examples will refer only
to the two inequivalent CP-conserving cases, namely, $e^{i\delta}=\pm1$.
In these two cases, the mixing matrix takes a real form $U_\mathrm{CP}$,
\begin{equation}
\label{UCP}
U_\mathrm{CP}=\left(
\begin{array}{ccc}
c_{13} c_{12} & s_{12}c_{13} & \pm s_{13}\\
-s_{12}c_{23}\mp s_{23}s_{13}c_{12} & c_{23}c_{12}\mp s_{23}s_{13}s_{12} &
s_{23}c_{13}\\
s_{23}s_{12}\mp s_{13}c_{23}c_{12} & -s_{23}c_{12}\mp s_{13}s_{12}c_{23}
& c_{23}c_{13}
\end{array}
\right)\ ,
\end{equation}
where the upper (lower) sign refers to $\delta=0$ ($\delta=\pi$). The
two cases are formally related by the replacement $s_{13}\to-s_{13}$.

The three-neutrino mass
spectrum $\{m_i\}_{i=1,2,3}$ is formed by a ``doublet'' of relatively close
states and by a third ``lone'' neutrino state, which may be either
heavier than the doublet (``normal hierarchy,'' NH) or lighter
(``inverted hierarchy,'' IH). Here, the lightest (heaviest) neutrino in the
doublet is called $\nu_1$ ($\nu_2$), so that their squared mass
difference is
\begin{equation}
\delta m^2=m^2_2-m^2_1>0
\end{equation}
by convention. The lone state is then labeled as $\nu_3$, and the physical
sign of $m^2_3-m^2_{1,2}$ distinguishes NH from IH.

Concerning the second independent squared mass difference $\Delta m^2$,
we define it as \cite{Ours,Qave}
\begin{equation}
\Delta m^2=\left| m^2_3-\frac{m^2_1+m^2_2}{2}\right|\ ,
\end{equation}
so that the two hierarchies (NH and IH) are simply related by the transformation
$+\Delta m^2\to-\Delta m^2$. The largest and next-to-largest squared
mass gaps are given $\Delta m^2\pm\delta m^2/2$ in both cases.
More precisely, the squared mass matrix $M^2$ reads, in such convention,
\begin{equation}
M^2=\mathrm{diag}(m^2_1,m^2_2,m^2_3)
=\frac{m^2_2+m^2_1}{2}\,\mathbf{1}+\mathrm{diag}
\left(
-\frac{\delta m^2}{2},+\frac{\delta m^2}{2},\pm\Delta m^2
\right)\ ,
\end{equation}
where the upper (lower) sign refers to normal (inverted) hierarchy.

In the previous equation, the term proportional to the unit matrix
$\mathbf{1}$ is irrelevant in neutrino oscillations, while it
matters in observables sensitive to the absolute neutrino mass
scale, such as in $\beta$-decay and precision cosmology. In
particular, $\beta$-decay experiments are sensitive
to the so-called effective electron neutrino mass $m_\beta$,
\begin{equation}
 \label{mb} m_\beta = \left[\sum_i|U_{ei}|^2m^2_i\right]^\frac{1}{2}=
\left[c^2_{13}c^2_{12}m^2_1+c^2_{13}s^2_{12}m^2_2+s^2_{13}m^2_3
\right]^\frac{1}{2}\ ,
\end{equation}
as far as the single $\nu_i$ mass states are not experimentally
resolvable. On the other hand, precision cosmology is
sensitive, to a good approximation (up to small 
hierarchy-dependent effects which may become important
in next-generation precision measurements \cite{Lesg})
to the sum of neutrino masses $\Sigma$ \cite{Past},
\begin{equation}
\Sigma = m_1+m_2+m_3\ .
\end{equation}

Finally, if neutrinos are indistinguishable from their antiparticles (i.e.,
if they are Majorana rather than Dirac neutrinos), the mixing
matrix $U$ acquires a (diagonal) extra factor 
\begin{equation}
U\to U\cdot U_M\ ,
\end{equation}
containing Majorana phases $\phi_i$,
which are irrelevant in oscillations but not in neutrinoless
double beta decay ($0\nu 2\beta$).  Using the parametrization
\begin{equation}
U_M=\mathrm{diag\left(1,e^{\frac{i}{2}\phi_2},
e^{\frac{i}{2}(\phi_3+2\delta)}\right)}\ ,
\end{equation}
the expression of the effective Majorana mass $m_{\beta\beta}$ 
probed in 
$0\nu2\beta$ experiments \cite{PDG4} takes the form:
\begin{equation}\label{mbb}
m_{\beta\beta} = \left|\sum_i U_{ei}^2 m_i\right| =\left|
c^2_{13}c^2_{12}m_1+c^2_{13}s^2_{12}m_2e^{i\phi_2}+s^2_{13}m_3
e^{i\phi_3}\right|\ .
\end{equation}

Finally, we remark that 
the constraints on the neutrino oscillation parameters shown hereafter
have been obtained by fitting accurate theoretical predictions
to a large set of experimental data, through either
least-square or maximum-likelihood methods. In both cases, parameter
estimations reduce to finding the minimum of a $\chi^2$ function
and to tracing iso-$\Delta \chi^2$ contours around it.
We adopt the convention used in \cite{PDG4} and call ``region
allowed at $n\sigma$'' the subset of the parameter space obeying
the inequality
\begin{equation}
\Delta \chi^2 \leq n^2 \ .
\end{equation}
The projection of such allowed region onto each single parameter
provides the $n\sigma$ bound on such parameter. In particular,
we shall also directly use the relation $\sqrt{\Delta\chi^2}=n$
to derive allowed parameter ranges at $n$ standard deviations.
Numerical results and figures are taken from the recent review 
\cite{Ours} (to which the reader is referred for details), 
which also makes use of results from Refs.~\cite{GeoR,Melc}.
For previous reviews on neutrino parameters see, e.g., 
Refs.~\cite{Focu,Mass,Conc,Barg,McDo,Gosw,Smir,Stru}.

\section{Constraints on $(\delta m^2,\sin^2\theta_{12},\sin^2\theta_{13})$
from solar+KamLAND  data}

It is well known that the angle $\theta_{13}$ is relatively small
and possibly zero. For $\theta_{13}=0$, both solar and
(long-baseline) reactor neutrino oscillations
depend solely on the parameters $(\delta m^2,\theta_{12})$.
Figure~1 shows the current constraints on such parameters from
a global analysis of all the available solar neutrino data \cite{Ours}
and of KamLAND data \cite{GeoR}, both separately and in combination. 
Although (at $3\sigma)$
multiple solutions can explain KamLAND data, the combination 
with solar data provides
a well-defined and unique solution at large mixing angle (LMA)
in the mass-mixing parameter space. The identification of
such solution represents one of the most impressive recent advances in
neutrino physics. 

Further progress can be expected in narrowing the
parameter space 
in Fig.~1. The $\delta m^2$ uncertainty is currently dominated by
the KamLAND observation of half-period of oscillations~\cite{Kam2}
and can be improved with higher statistics \cite{Berg}. The
$\sin^2\theta_{12}$ uncertainty is instead dominated by the
SNO ratio of charged-to-neutral current (CC/NC) event rates, which
can also be improved with future data \cite{Mikn}.

The current solar LMA solution, as
compared with results prior to the complete
SNO-II data set \cite{Mikn}, 
is slightly shifted toward larger
values of $\sin^2\theta_{12}$ and allows higher values of $\delta
m^2$. [Our current best-fit point for solar data only is at $\delta
m^2=6.3\times 10^{-5}\mathrm{\ eV}^2$ and $\sin^2\theta_{12}=0.314.$]
This trend is substantially due to the larger value of the CC/NC
ratio measured in the complete SNO II phase (0.34 \cite{SNOL,Mikn}) 
with respect to the previous
central value (0.31 \cite{SNO2}). 
We also find that the SNO-II charged-current spectral data \cite{SNOL} 
contribute to allow slightly higher values of $\delta m^2$ with
respect to older results.

For $\theta_{13}>0$, the solar and KamLAND $\nu$ parameter space is
spanned by $(\delta m^2,\sin^2\theta_{12},\sin^2\theta_{13})$. 
Figure~2 shows the current $2\sigma$ bounds in such space (both separately and in
combination) in each of the three coordinate planes. 
Remarkably, both solar and KamLAND data are consistent
with $\theta_{13}$ being small ($\sim 0$ at best-fit), in agreement with
independent atmospheric, accelerator and short-baseline reactor
data (see the next section). The combined upper bound on
$\sin^2\theta_{13}$ in Fig.~2 is at the interesting
level of $\sim 5\%$.

\section{Constraints on $(\Delta m^2,\sin^2\theta_{23},\sin^2\theta_{13})$
from SK$_\mathrm{ATM}$+K2K+CHOOZ  data}

In the limit $\delta m^2/\Delta m^2\ll 1$ (one-dominant-mass-scale
approximation), the leading parameters in atmospheric and long-baseline
accelerator searches are $(\Delta m^2,\sin^2\theta_{23},\sin^2\theta_{13})$.
Subleading effects induced by $\delta m^2\neq 0$ (i.e., LMA
effects in terrestrial neutrino oscillations \cite{Pere,Mal1,RCCN}) 
are present, however,
even for $\theta_{13}\neq 0$ \cite{Mal2}. In accurate calculations, it
is worthwhile to include such effects numerically, e.g., by fixing
($\delta m^2,\sin^2\theta_{12}$) at their best-fit value in Fig.~1
in a full three-flavor analysis of atmospheric and K2K data, as done
in the following two figures. 

Figure~3 shows, for $\theta_{13}=0$, 
the results of our analysis of SK (atmospheric) and K2K data,
both separately and in combination. 
The K2K constraints are octant-symmetric and relatively
weak in $\sin^2\theta_{23}$, while they contribute appreciably to
reduce the overall $\Delta m^2$ uncertainty. The SK atmospheric
neutrino contraints are instead strong on both mass and mixing
parameters, and also slightly asymmetrical \cite{Mal2} in $\sin^2\theta_{23}$.
Unfortunately, current data are not accurate enough to promote
this slight asymmetry to a real $\theta_{23}$-octant discrimination.
However, it 
is not excluded that future, high-statistics atmospheric neutrino data
might be able to do so, if $\theta_{23}$ is not too close to
$\pi/4$ \cite{Mal3}. Such possible $\theta_{23}$ 
octant asymmetry, together with
a measurement of $\theta_{13}$, is crucial for model building
\cite{Tani,Frig}.

For $\theta_{13}>0$, SK+K2K data are also sensitive, in principle,
to the neutrino mass hierarchy [$\mathrm{sign}(\pm\Delta m^2)=\pm 1$] and to the
CP parity [$\cos\delta=\pm1$]. However, the dependence is very small within
the CHOOZ bounds on $\theta_{13}$
(see, e.g., Ref.~\cite{Ours} and references therein), 
and thus it makes sense to marginalize the
SK+K2K+CHOOZ $\chi^2$ function with
respect to hierarchy and CP parity. The results are shown in
Fig.~4, in terms of the projections of the $(\Delta
m^2,\sin^2\theta_{23},\sin^2\theta_{13})$ region allowed at 1, 2,
and $3\sigma$ onto each of the coordinate planes (with LMA effects
included). The best fit is
reached for nonzero $\theta_{13}$ (mainly due to a slight preference of
low-energy atmospheric data for $\nu_e$ event appearance), but
$\theta_{13}=0$ is allowed within less than $1\sigma$.  The
preferred value of $\sin^2\theta_{23}$ remains slightly below
maximal mixing. The best-fit value of $\Delta m^2$ is $2.4\times
10^{-3}$ eV$^2$. Notice that the correlations among the three
parameters in Fig.~4 are very weak.

\section{Global constraints on oscillation parameters}

The results of the global analysis of solar+KamLAND data (Sec.~2) and of SK+K2K+CHOOZ data (Sec.~3) can now be merged to
provide our best estimates of the five neutrino oscillation 
parameters $(\delta m^2,
\Delta m^2, \theta_{12}, \theta_{13}, \theta_{23})$, marginalized
over the $2\times 2$ cases with different mass hierarchies and
CP parities (which are physically different but phenomenologically
indistinguishable at present). The bounds will be
directly shown in terms of the ``number of sigmas'', corresponding
to the function $(\Delta\chi^2)^{1/2}$ for each parameter.

Figure~5 shows our global bounds on $\sin^2\theta_{13}$, as coming from
all data (solid line) and from the following partial data sets: KamLAND (dotted),
solar (dot-dashed), solar+KamLAND (short-dashed)
and SK+K2K+CHOOZ (long-dashed). Only the latter set,
as observed before, gives a weak indication for nonzero $\theta_{13}$.
Interestingly, solar+KamLAND data are now sufficiently accurate
to provide bounds which are not much weaker
than the dominant SK+K2K+CHOOZ ones, also because the latter
slightly prefer $\theta_{13}>0$ as best fit, while the former do not.

Figure~6 shows our global bounds on the four mass-mixing
parameters which present both upper and lower limits with
high statistical significance. Notice that the accuracy of the
parameter estimate is already good enough to lead to almost
``linear'' errors, especially for $\delta m^2$ and
$\sin^2\theta_{12}$. For $\Delta m^2$ and $\sin^2\theta_{23}$,
such ``linearity'' is somewhat worse in the region close to the
best fit (say, within $\pm 1\sigma$), and thus $2\sigma$ (or $3\sigma$)
errors should be taken as reference.

\section{Non-oscillation data and their interplay with oscillation 
constraints}

Since oscillation data fix the mass splittings $\delta m^2$ and $\Delta m^2$,
the observables sensitive to absolute neutrino masses 
$(m_\beta,m_{\beta\beta},\Sigma)$ are higly correlated with each other, both
in normal and in inverted hierarchy: typically, when one increases the
other also increases. Therefore, upper bounds on any of them translate
into upper bounds on the others. However, the upper bounds on $m_\beta$ 
are currently weak (a few eV) \cite{Eite}, and the relevant discussion can be
limited to $(m_{\beta\beta},\Sigma)$ at present.   

Figure~7 shows the impact of all the available non-oscillation
data, taken at face value, in the parameter space
$(m_{\beta\beta},\Sigma)$, at the $2\sigma$ level. The horizontal band is
allowed by the positive $0\nu2\beta$ experimental claim \cite{Kl04}
equipped with the nuclear uncertainties of \cite{Rodi} as
described in \cite{Ours}. The slanted bands (for normal and inverted
hierarchy) are allowed by all other neutrino data, i.e., by the
combination of neutrino oscillation constraints (from Figs. 5 and 6)
and of astrophysical and cosmological constraints
from Cosmic Microwave Background (CMB) \cite{Be03}, large scale structures 
from galaxy surveys (2dF) \cite{Gala}, and small scale structures from
Lyman $\alpha$ forest data \cite{Lyma}, as described in 
\cite{Melc}.  
The tight cosmological upper bound on $\Sigma$
prevents the overlap between the slanted and horizontal bands at
$2\sigma$, indicating that no global combination of oscillation
and non-oscillation data is possible in the sub-eV range. 
The ``discrepancy'' is now
even stronger than it was found in Ref.~\cite{Melc},
due to the adoption of smaller $0\nu2\beta$
nuclear uncertainties \cite{Rodi}. It is
premature, however, to derive any definite conclusion as to which
piece of the data (or of the $3\nu$ scenario) is ``wrong'' in this
conflicting picture. E.g., by relaxing either the $0\nu2\beta$
lower bound or the Ly$\alpha$ data, global combinations are possible
\cite{Ours}. Further experimental and theoretical research
is needed to clarify the interplay of
absolute neutrino observables in the sub-eV
range.

\section{Summary and Conclusions}

There is compelling evidence for neutrino flavor change driven
by nonzero masses and mixing angles. Basically all oscillation
data (with the only exception of LSND) are consistent within a three-neutrino
framework. Within such framework, the 
global constraints from oscillation data can be summarized
 (see also Figs.~5 and 6
and Ref.~\cite{Ours})
through the following $\pm 2\sigma$ ranges
(95\% C.L.) for each parameter:
\begin{eqnarray}
\sin^2\theta_{13} &=& 0.9^{+2.3}_{-0.9}\times 10^{-2}\ ,\\
\delta m^2 &=& 7.92\, (1\pm 0.09)\times 10^{-5}\mathrm{\ eV}^2\ ,\\
\sin^2\theta_{12} &=& 0.314 \,(1^{+0.18}_{-0.15})\ ,\\
\Delta m^2 &=& 2.4\,(1^{+0.21}_{-0.26})
\times 10^{-3}\mathrm{\ eV}^2\ ,\\
\sin^2\theta_{23} &=&0.44\,(1^{+0.41}_{-0.22})\ .
\end{eqnarray}
Such ranges are marginalized over the four inequivalent cases
$[\mathrm{sign}(\pm\Delta m^2)]\otimes[\cos\delta=\pm 1]$, i.e., over
the two possible hierarchies and the two possible CP-conserving cases,
which are currently undistinguishable. [Notice that the lower error on
$\sin^2\theta_{13}$ is purely formal, and corresponds to the
positivity constraints $\sin^2\theta_{13}>0$.]

Concerning the observables sensitive to absolute masses
($m_\beta$, $m_{\beta\beta}$ and $\Sigma$), the situation is still unclear.
Current constraints at the eV/sub-eV level
are dominated by either upper bounds on $\Sigma$
from cosmology or by the $0\nu2\beta$ claim on $m_{\beta\beta}$, whose
combination is not possible, however, at face value. Further studies 
and data are need to go beyond the general statement that neutrino masses
should be smaller than $\sim1$~eV, and to really explore the sub-eV
range.

Within the three-neutrino scenario, it appears that
the most important unsolved
problems require probing $\theta_{13}$, $\delta$, the hierarchy, and the absolute neutrino masses. Needless to say, further experimental results
or theoretical insights might also reserve big surprises and force
us to go beyond such scenario, either by adding new neutrino states, or
new interactions, or both.

\section*{Acknowledgments}
This work is supported by the
Italian Ministero dell'Istruzione, Universit\`a e Ricerca (MIUR) and
Istituto Nazionale di Fisica Nucleare (INFN) through the
``Astroparticle Physics''  project.

\begin{figure}[th]
\begin{center}
\vspace*{-0.0cm}
\hspace*{-0.0cm}
\epsfig{file=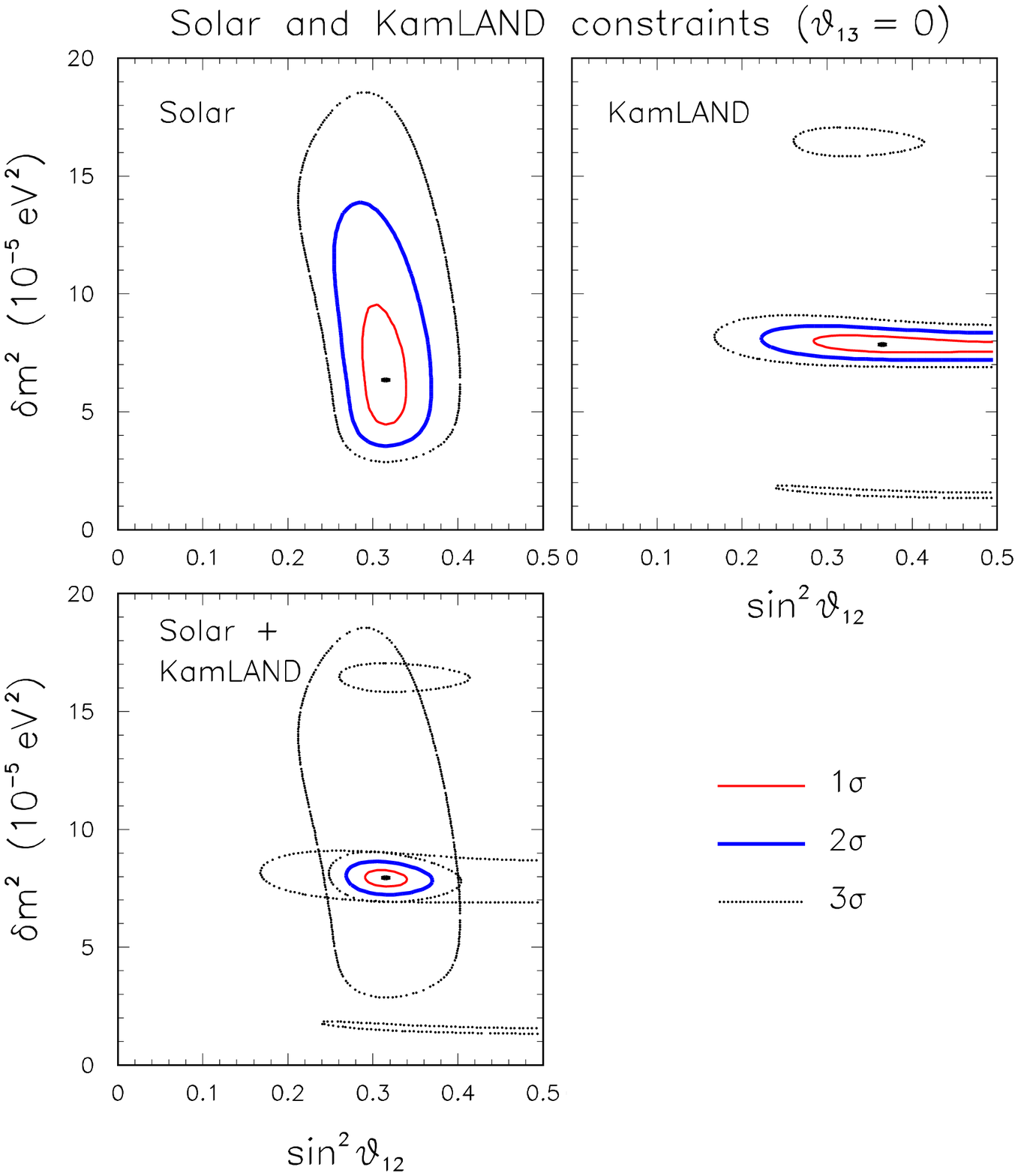,scale=0.54}
\caption{Solar and KamLAND constraints
in the mass-mixing plane $(\delta m^2,\sin^2\theta_{12})$
and for $\theta_{13}=0$, shown both separately and in combination,
at 1, 2, and $3\sigma$ level.
\label{fig_01}}
\vspace*{+1.2cm}
\hspace*{-0.0cm}
\epsfig{file=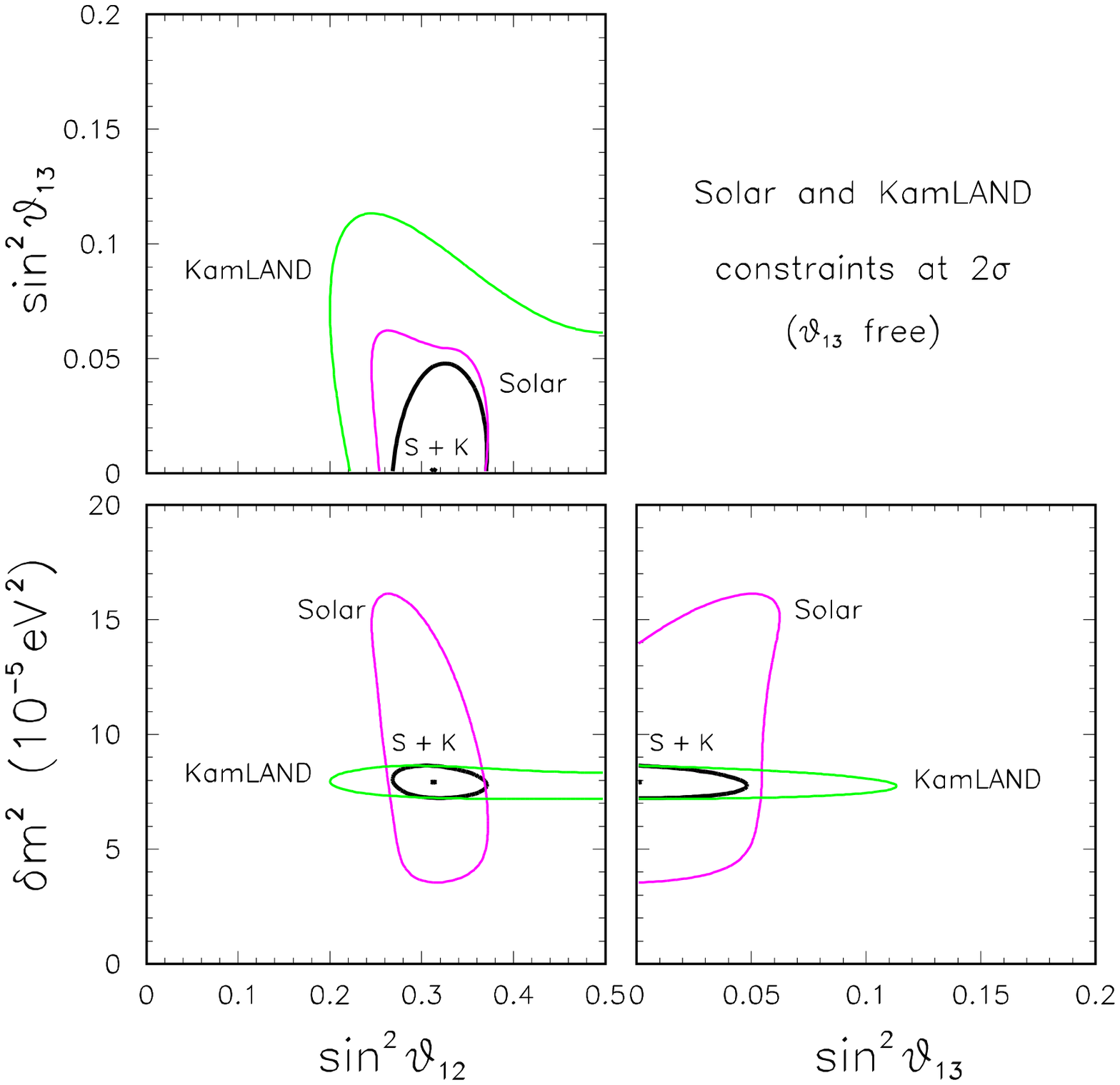,scale=0.54}
\caption{Three flavor analysis of solar and KamLAND data (both
separately and in combination) in the parameter space $(\delta
m^2,\sin^2\theta_{12},\sin^2\theta_{13})$. 
The contours represent projections of
the region allowed at $2\sigma$ ($\Delta\chi^2 = 4$).
\label{fig_02}}
\end{center}
\end{figure}

\begin{figure}[th]
\begin{center}
\vspace*{-0.0cm}
\hspace*{-0.0cm}
\epsfig{file=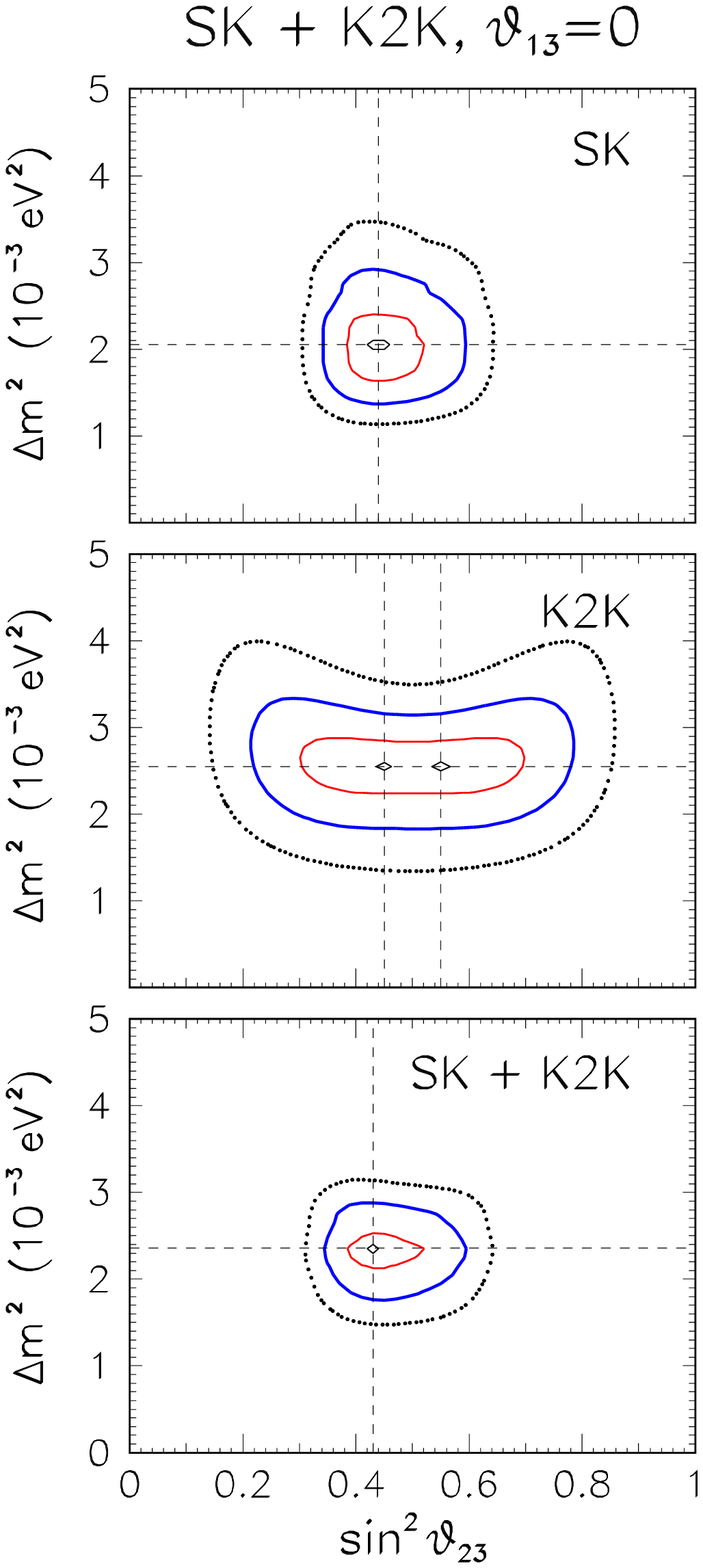,scale=0.47}
\caption{Analysis of SK and K2K data (both separately and in
combination) in the plane $(\Delta m^2,\sin^2\theta_{23})$ at
$\theta_{13}=0$. The parameters $(\delta m^2,\sin^2\theta_{12})$
have been fixed at their best-fit LMA values.
\label{fig_03}}
\vspace*{+1.2cm}
\hspace*{-0.0cm}
\epsfig{file=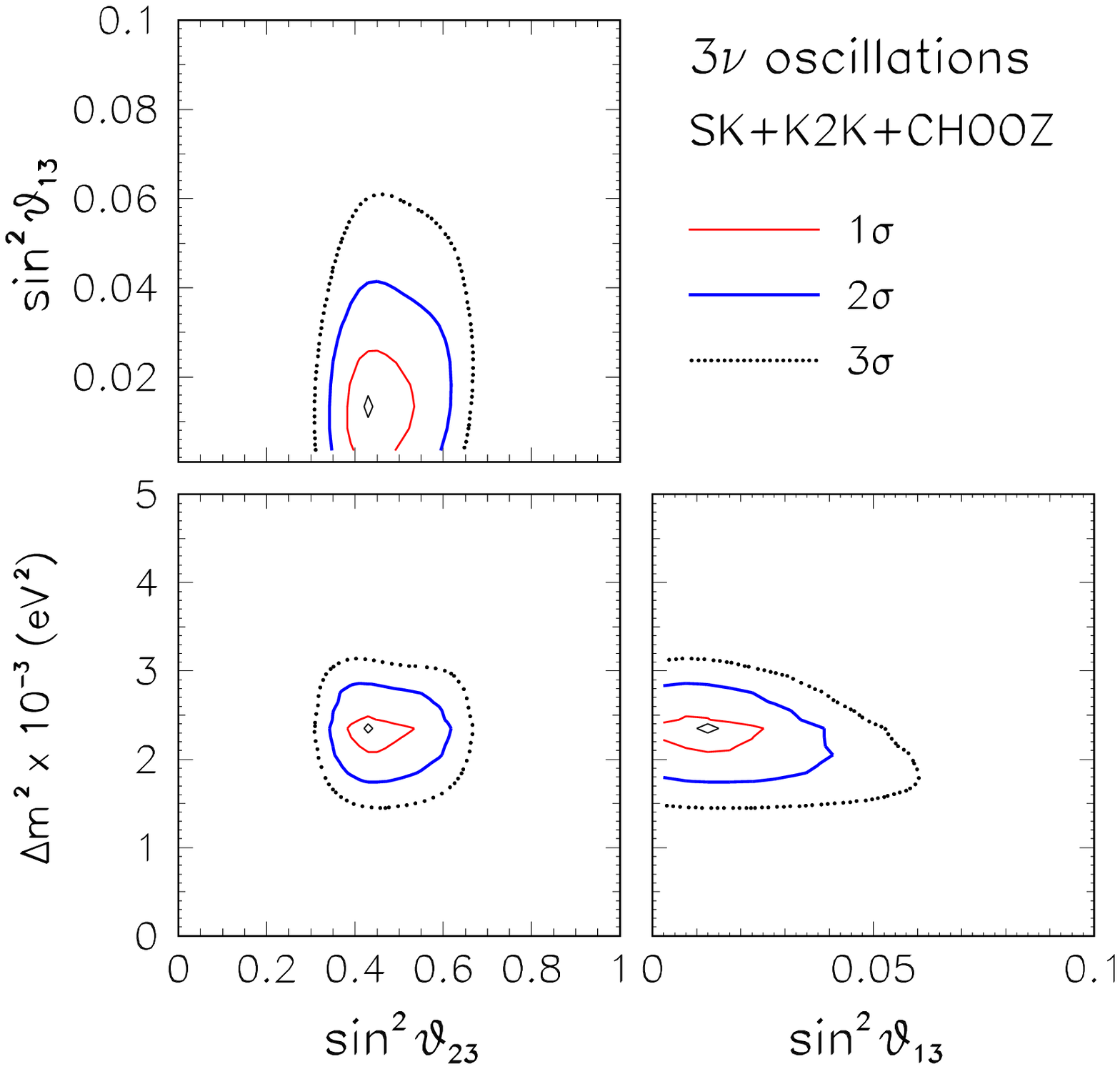,scale=0.54}
\caption{Three-neutrino analysis of SK+K2K+CHOOZ data, including
subleading LMA effects. The results are shown as
projections of the
$(\Delta m^2,s^2_{23},s^2_{13})$ allowed regions (at 1, 2, and $3\sigma$),
marginalized with respect to the four cases
$[\cos\delta=\pm 1]\otimes[\mathrm{sign}(\pm \Delta m^2)=\pm 1]$.
\label{fig_04}}
\end{center}
\end{figure}

\begin{figure}[th]
\begin{center}
\vspace*{-0.0cm}
\hspace*{-0.0cm}
\epsfig{file=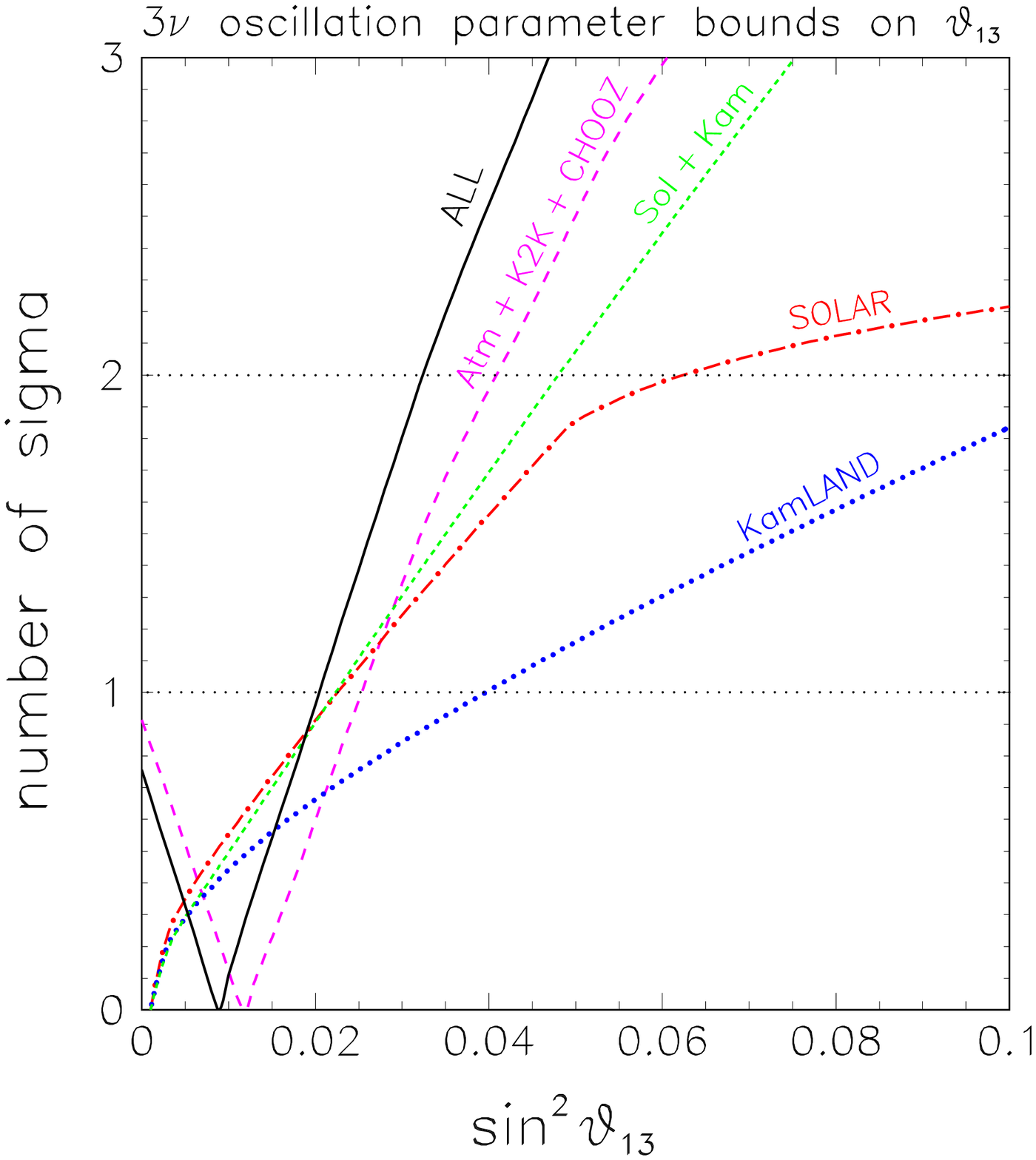,scale=0.4}
\caption{Global three-neutrino analysis of oscillation data. Bounds
on $s^2_{13}$ are shown in terms of $n\sigma=\sqrt{\Delta\chi^2}$ for
 KamLAND (dotted curve), solar (dot-dashed curve), solar+KamLAND
(short-dashed curve), SK+K2K+CHOOZ (long-dashed curve) and all
data combined (solid curve). In each case, the continuous
parameters $(\Delta m^2,s^2_{23},s^2_{13})$ and---if
applicable---the discrete parameters $[\cos\delta=\pm
1]\otimes[\mathrm{sign}(\pm \Delta m^2)=\pm 1]$ are marginalized
away.
\label{fig_05}}
\vspace*{+1.2cm}
\hspace*{-0.0cm}
\epsfig{file=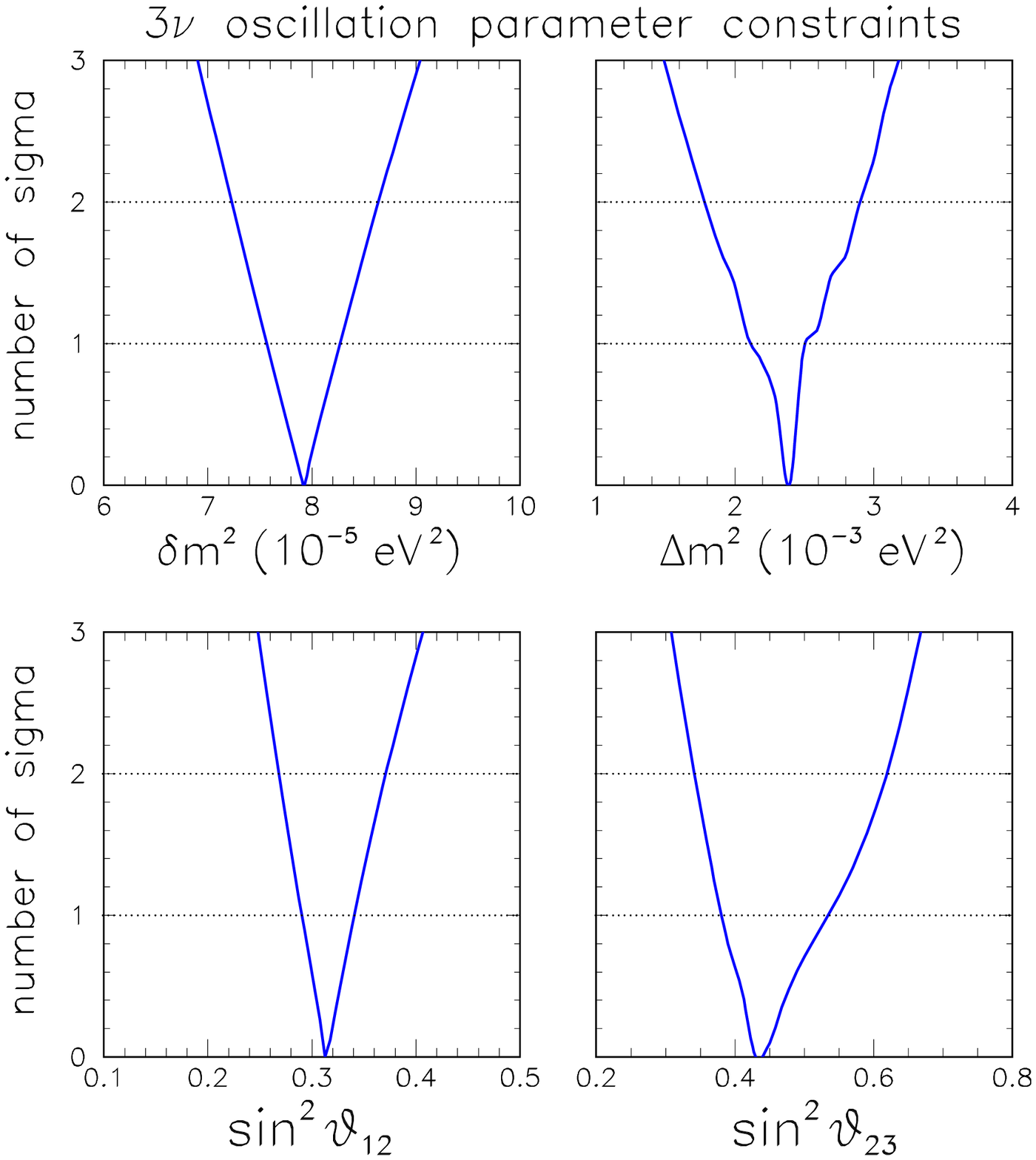,scale=0.54}
\caption{Global three-neutrino analysis of  oscillation data.
Bounds on each of the parameters $\delta m^2$, $\Delta m^2$,
$\sin^2\theta_{12}$, and $\sin^2\theta_{23}$ are shown in terms of
$n\sigma=\sqrt{\Delta\chi^2}$. In each plot, all parameters but
the one in abscissa are marginalized away.
\label{fig_06}}
\end{center}
\end{figure}

\begin{figure}[t]
\begin{center}
\vspace*{+0.0cm}
\hspace*{-1.2cm}
\epsfig{file=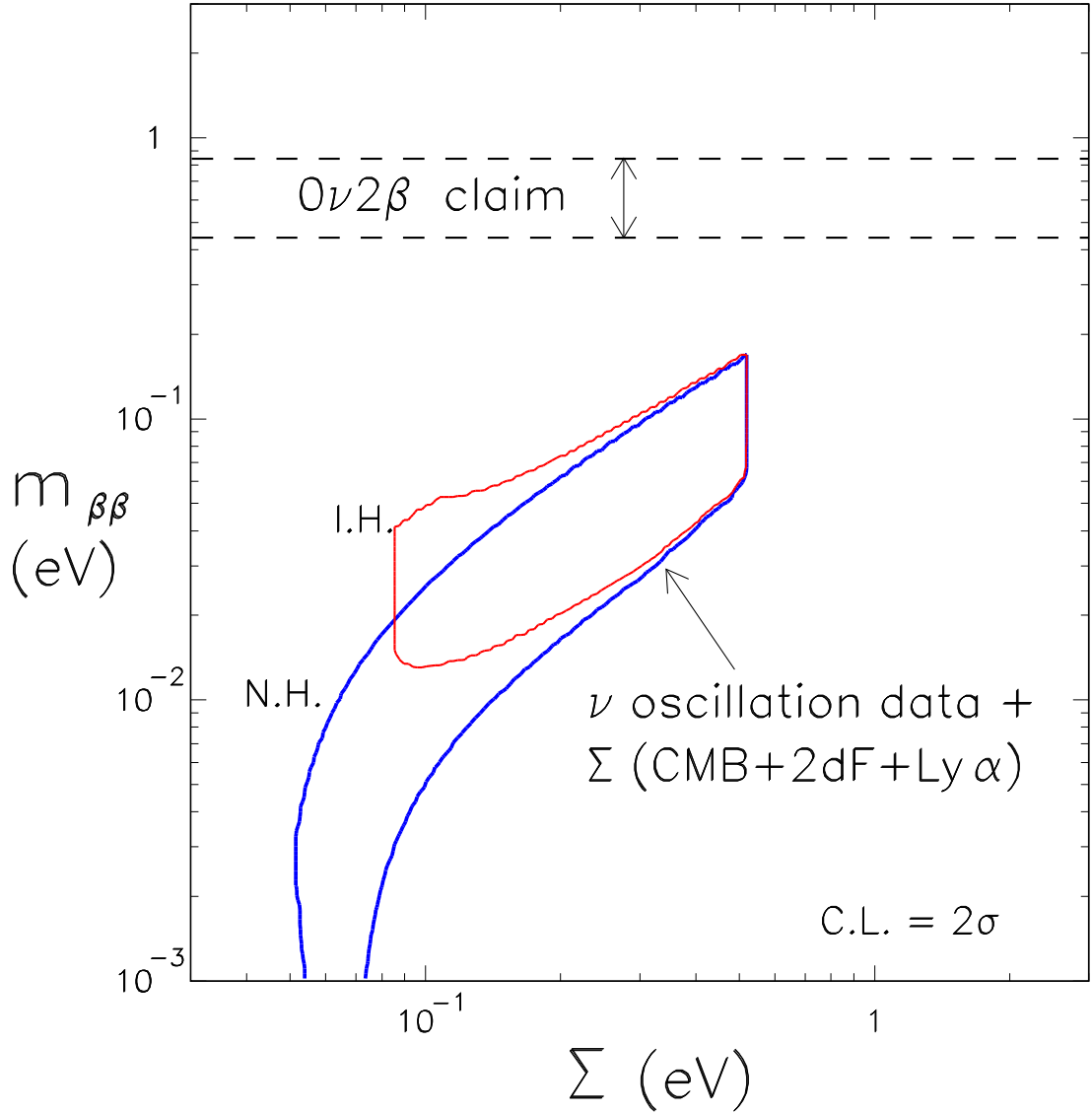,scale=0.8}
\caption{Analysis of oscillatory and non-oscillatory observables
in the plane $(m_{\beta\beta},\Sigma)$. The $2\sigma$ horizontal
band is preferred by the positive $0\nu2\beta$ claim, while the
slanted $2\sigma$ regions below (for normal hierarchy, N.H.,
and inverted hierarchy, I.H.) are preferred by all other data
(i.e., oscillation plus cosmological data). The absence of overlap
indicates tension among the data in the sub-eV range, as far as
the standard three-neutrino framework is assumed. See 
also Refs.~\protect\cite{Ours,Melc}
\label{fig_07}}
\end{center}
\end{figure}

\end{document}